\documentclass[twocolumn,showpacs,preprintnumbers,amsmath,amssymb]{revtex4}

\usepackage{float}
\usepackage{graphicx}
\usepackage{dcolumn}
\usepackage{bm}
\usepackage[dvips]{color}
\definecolor{grau}{rgb}{0.9,0.9,0.9}
\definecolor{hellgrau}{rgb}{0.935,0.935,0.935}
\definecolor{gelb}{rgb}{1,1,0.7}
\definecolor{rot}{rgb}{1,0.875,0.935}
\definecolor{blau}{rgb}{0.1,0.21,0.5}
\definecolor{gruen}{rgb}{0.25,0.5,0.35}
\definecolor{hellgelb}{rgb}{0.4,0.5,0.5}
\definecolor{dgelb}{rgb}{0.6,0.5,0.5}
\definecolor{dunkelgruen}{rgb}{0.0,0.45,0.0}
\definecolor{dunkelrot}{rgb}{0.5,0.06,0.05}
\definecolor{dt}{rgb}{0.75,0.65,1}
\definecolor{tuerkis}{rgb}{0.95,0.9,1}

\begin{document}

\preprint{APS/123-QED}

\title{The effect of the quantization of the centrifugal stretching on the analysis of rotational spectra of even-even deformed nuclei in rare earth and actinide regions}

\author{Abdurahim A. Okhunov }
 \email{aaokhunov@gmail.com}
\affiliation{Namangan Institute of Engineering and Technology, 160115 Namangan, Uzbekistan}%

\affiliation{ Department of Science in Engineering, KOE,
International Islamic University Malaysia, P.O Box 10, 50728
Kuala Lumpur, Malaysia }%



\author{Mohd Kh. M. Abu El Sheikh}%
 \email{Corresponding author: mhsr70@gmail.com}
\affiliation{Department of Physics, University of Malaya, 50603
Kuala Lumpur, Malaysia }%



\date{\today}

\begin{abstract}
Deviation from the $I\left(I+1\right)$ rule in the even–even isotopes $_{62}Sm$, $_{64}Gd$, $_{66}Dy$, $_{68}Er$, $_{70}Yb$,
$_{72}Hf$, $_{74}W$ and $_{76}Os$ nuclei have been studied with a new approach based on the idea that the rotational nucleus
being stretched out along the symmetry axis, where such stretching is treated according to quantum mechanics. This approach
led to a new formula that describes the dependence of the moment of inertia on the angular momentum, and such formula works
well for all even-even nuclei in the actinide and rare-earth region in the range $2.9<R=\frac{E_4}{E_2}<3.33$. The calculations
were carried out with the stretched model where the stretching of the nucleus with the rotation is quantized. The obtained
results with a new derived formula which gives a reasonable agreement with the experimental date for all I up to 16.


\end{abstract}



\maketitle

\section{\label{sec:level1}Introduction }


Bohr in his model of "Coupling between the motion of the single particle to the nuclear surface oscillation"
predicted that, in the special case of the strong-coupling where the nucleus is well deformed, the nuclear
spectrum is rotational with spin sequence $I=0,2,4,6,...$, even parity and with energy levels follows the
simple formula ~\cite{Bohr51,Bohr52,Bohr76}.


\begin{eqnarray}\label{Eq.1}
E(I)=\frac{\hbar^{2}}{2\Im}I(I+1),
\end{eqnarray}
\noindent
where $\Im$ is the moment of inertia which is expected to be constant for such nuclei. According
to such model, nuclei having these characteristics are expected to be in the region where a large number of
particles are outside the closed shell. However, it was shown later that the energy spacing experimentally
measured of the states of the ground band increases less rapidly than that is predicted by the formula
~Eq. (\ref{Eq.1}) when $\Im$ is constant. Originally, Bohr and Mottelson ~\cite{Bohr53} suggested that this
decrease in the energy spacing may be understood within the context of the coupling between rotational and
vibrational modes of motion which contributes a term of the form $-B\left[I\left(I+1\right)\right]^2$, and
~Eq. (\ref{Eq.1}) becomes

\begin{eqnarray}\label{Eq.2}
E(I)=AI(I+1)-B\left[I\left(I+1\right)\right]^{2}
\end{eqnarray}

The two parameters $A$ and $B$ in this equation are
related to the moment of inertia and the rotation-vibration coefficient respectively. However, such correction was found
to be insufficient to represent the experimental spectra obtained with high spin $I$- values or even
for low lying states of transitional nuclei \cite{Preston018}. One can generalize ~Eq. (\ref{Eq.2}) to a
power series form \cite{Sood67}

\begin{eqnarray}\label{Eq.3}
E(I)&=&AI(I+1)-B\left[I\left(I+1\right)\right]^{2} \\ \nonumber
&+&C\left[I\left(I+1\right)\right]^{3}-D\left[I\left(I+1\right)\right]^{4}+... ,
\end{eqnarray}

\noindent
where $A, B, C, D,...$ are parameters, which is can be determined by fitting this equation with experimental results.
The problem here is that, for a feasible comparison, we require as many parameters as there are experimental data to
be fitted. Gupta \cite{Gupta67, Gupta69, Trainor71} introduces a model with a rotational-vibrational interaction that
relates all parameters $C, D, E,...$ to the first two basic parameters $A, B$. This model is known as non-rigid rotator
model and the energy states are expressed as a power series in terms of $B/A$ and the weight factor $I(I+1)$ as:

\begin{eqnarray}\label{Eq.4}
E(I)&=&AI(I+1)\left[1-\frac{A}{B}I\left(I+1\right)\right. \\ \nonumber
&&\hspace*{0.9cm}+\left.3\left(\frac{A}{B}I\left(I+1\right)\right)^2-...\right].
\end{eqnarray}

It has been shown that the first three terms and even the first four terms of ~Eq. (\ref{Eq.4}) are capable
of solving the problem only within a very short range $3.25<R<3.33$, Figure 1.	

However, a detailed analysis by Sood \cite{Sood67, Sood68, Volkov71} showed that the using of any truncated series in $I(I+1)$
in ~Eq. (\ref{Eq.4}) is not sufficient to describe the experimental spectrum. Instead he suggested that the phenomenological sum an
infinite series in $I(I+1)$ to gate the following compact expression for the energy $E(I)$

\begin{eqnarray}\label{Eq.5}
E(I)=A(I)I(I+1),
\end{eqnarray}

\noindent where $A(I)=A\left[1-\frac{\frac{B\cdot I(I+1)}{A}}{1+\frac{NB}{A}\cdot I(I+1)}\right]$,
which could be consistent with that derived by Bohr for describing the experimental spectrum with the
provision that $A(I)$ in ~Eq. (\ref{Eq.5}) is variable and it is a decreasing function of  $I$.
Although this two-parameter model by Sood \cite{Sood67} fitted with high accuracy the experimental spectrum
of rare-earth nuclei including the $Os$ isotopes and $N=90$ nuclei, there are many questions about the parameter
$N$ has been left open such as; what is the physical quantity that the parameter $N$  in ~Eq. (\ref{Eq.5}) represents?
Or what are the factors that $N$ depends on? The answer to these questions is very important to specify the
factors which control the increase of $\Im$ with the increase in $I$.

An alternative approach to interpret such decreasing in the energy space was introduced first by Morinaga \cite{Morinaga66}
and later by Diamond, Stephen and Swiatecki \cite{Diamond64}. Morinaga \cite{Morinaga66} was the first who
suggested that the decrease in energy spacing is due to the increase in the moment of inertia $\Im$.
Diamond et al. ~\cite{Diamond64}, attribute this increase in $\Im$ to some sort of centrifugal stretching.
This model was known as the beta-stretching model. Let us assume the suggestion of Bohr that the total energy
$E_I$ of the state $I$ is defined as

\begin{eqnarray}\label{Eq.6}
E(I)=\frac{\hbar^{2}}{2\Im}I(I+1)+\frac{1}{2}C\left(\beta_{I}-\beta_{0}\right)^{2},
\end{eqnarray}

\noindent
where $C$ is the stiffness of the nucleus. The authors in reference ~\cite{Diamond64} assume the hydrodynamical formula $\Im\infty\beta^{2}$
for the moment of inertia. Where $\beta_{I}$ in ~Eq. (\ref{Eq.6}) is the value of $\beta$ which satisfy the condition

\begin{eqnarray}\label{Eq.7}
\frac{\partial E(I)}{\partial\beta}=0,
\end{eqnarray}

\noindent
so that the total energy is minimized. This approach gives good fit only for strongly deformed nuclei. Bands outside this region cannot be fitted with reasonable accuracy.

M.A.J. Mariscotti et al. \cite{Mariscotti69} thought that the moment of inertia $\Im$ depends, not only on the deformation
parameter $\beta$ but also on Coriolis pairing effect, therefore, they suggested $\beta$ in ~Eq. (\ref{Eq.7}) should
be replaced by a general variable $t$ which might represents not only the deformation parameter, but also all other
microscopic features like the effective pairing. The dependence of $\Im$ from $t$ can be expressed as
$\Im=const\cdot t^{n}$, $n$ being integer. Since the best fit for all ground state bands ranging $2.34\leq R\leq 3.33$
were obtained at $n=1$, then $\Im$ itself, can be considered as a general variable and the equation of the total energy
for the ground state bands, ~Eq. (\ref{Eq.7}), takes the form,

\begin{eqnarray}\label{Eq.8}
E(I)=\frac{\hbar^{2}}{2\Im}I(I+1)+\frac{1}{2}C\left(\Im_{I}-\Im_{0}\right)^{2},
\end{eqnarray}

\noindent with the equilibrium condition

\begin{eqnarray}\label{Eq.9}
\frac{\partial E(I)}{\partial\Im}=0,
\end{eqnarray}

\noindent
must be satisfied for each state of $I$.
The model of Mariscotti and et. al. \cite{Mariscotti69} is known as the variable moment of inertia ($VMI$) model. In spite of
the great success of this model in getting excellent fit for all ground-state bands ranging $2.34\leq R\leq 3.33$,
but, like Sood, the authors considered the moment of inertia as a general variable and they did not clarify
precisely how this general variable depends on the deformation parameter and the microscopic features and also
what is the ratio that each parameter contribute.

Harris suggested an expansion of both; total energy and the moment of inertia in power of $\omega^2$ instead of
$I\left(I+1\right)$ expansion of Bohr and Mottelson, ~Eq, (\ref{Eq.3}). It was found excellent agreement with
energies in the ground state bands of deformed nuclei \cite{Harris65}.
In reference \cite{Okhunov015}, a theoretical analysis of the deviation from the adiabatic theory (i.e the case
of strong coupling mentioned before) was presented on the basis of phenomenological model ($Phen.M$)\cite{Usmanov010},
take into account the Coriolis mixture of low-lying state bands. Authors \cite{Okhunov015} calculated the values of moment of inertia
$\Im_{0}$, $\Im_{1}$ using Harris parametrization \cite{Harris65}, and energy spectra of rotational ground ($gr$)
state band of even-even $^{152-156}Sm$, $^{156-160}Gd$, $^{156-166}Dy$, $^{166-176}Yb$, $^{170-180}Hf$ and
$^{174-184}W$ nuclei. The obtained results show a very good agreement with experimental data.
Furthermore, recently for the even-even deformed $^{154,156}Sm$, $^{156-160}Gd$, $^{160-164}Dy$, $^{164-170}Er$,
$^{170-176}Yb$, $^{176-180}Hf$, $^{182-186}W$, $^{186-190}Os$, $^{194-198}Pt$, $^{222-228}Ra$, $^{226-234}Th$,
$^{230-238}U$ and $^{238,240}Pu$ nuclei the moment of inertia has been calculated by the theoretical
extension of hydrodynamical model in view of the contributions arising from higher order terms of radial
distribution \cite{Muhammad020}. Such calculated values are found to be in better agreement than the original model for both
axially deformed and triaxial nuclei. This highlights the crucial approximation involved in the irrotational
picture of liquid droplet in terms of small amplitude vibrations and further supports the large amplitude vibrations
at the nuclear surface.

A nuclear rotation-vibration model based on the cranking Bohr-Mottelson Hamiltonian has been successfully applied
to normal rotational bands of only the well deformed even-even nuclei in reference ~\cite{Hu93}.
Recently, a three parameters formula were derived by H. X. Huang, C. S. Wu, and J. Y. Zeng ~\cite{Huang89}.
This formula was analyzed the rotational spectra up to $I=20$. It shows an excellent agreement with experiment
for all actinide and rare earth nuclei only in well-deformed region i.e. $3.2<R<3.33$. Like Diamond and Gupta,
this formula doesn’t work well for nuclei outside of the strongly deformed region.

The peresent work aimed to show that the idea of a spinning nucleus being stretched out along the symmetry
axis that is mentioned in reference ~\cite{Diamond64}, when such stretching is quantized, leads to the derivation
of a formula for the moment of inertia that is capable to reproduce level energies in the ground state bands up to
$I=16$ for all actinide and rare earth nuclei in well-deformed as well as in transitional regions in the range $2.9<R<3.33$.
As a consequence, one can prove that the moment of inertia depends mainly on the deformation parameter of the nucleus.
The results of this work showed that the effect of all microscopic features like Coriolis forces and pairing effect is
so small and it can be neglected in many cases. A brief description of the formulation of this formula will be displayed
in section (2). Application of this formula to the ground state bands of an even-even nuclei whose energy ratio not less
than 2.9 is given in section 3. In section 4 the important conclusions will be summarized.

\section{\label{sec:level2}Formalism of the Model}

In the present model, we shall follow the suggestion of Diamond et al. \cite{Diamond64}, that the increase
in the moment of inertia can be interpreted on the basis of the idea that spinning nucleus exhibits some sort of
centrifugal stretching along the symmetric axis, the very simple classical form of the moment of inertia of the mass element $dm$ is

\begin{eqnarray}\label{Eq.10}
d\Im_{I}=r^{2}dm,
\end{eqnarray}

\noindent
where $r$ is the effective radius of rotation. According to the hydrodynamical model, a very little of the
nuclear matter is actually taking part in the effective rotational motion or, in other words, the rotational
motion can be pictured as a motion of wave around the nuclear surface, so

\begin{eqnarray}\label{Eq.11}
dm=\frac{M}{A}R^{2}d\Omega=\frac{M}{4\pi}d\Omega,
\end{eqnarray}

\noindent
where $A\approx4\pi R^2$ is the total area of the surface, $R^2d\Omega\approx R^2\sin\theta d\theta d\phi$ is the
surface element, $R$ is the distance of the mass element $dm$ from the center of the nucleus. Where $r$ in ~Eq, (\ref{Eq.10})(10)
can be written as $r_{0}+\bigtriangleup r$ where $\bigtriangleup r$ is the stretched in the nucleus due to rotation. It follows

\begin{eqnarray}\label{Eq.12}
d\Im=\left(r_{0}+\triangle r\right)^{2}\frac{M}{4\pi}d\Omega=r_{0}^{2}\left(1+\frac{\triangle r}{r_{0}}\right)^{2}\frac{M}{4\pi}d\Omega .
\end{eqnarray}

We assume that the nucleus has symmetry axis which is perpendicular to its rotational axis then the stretched can be expanded in a complete set of spherical harmonics as $\frac{\triangle r}r_{0}=\sum_{\lambda\mu}a_{\lambda\mu}^{*}Y_{\lambda\mu}\left(\theta,\phi\right)$.
Putting this definition in to ~Eq. (\ref{Eq.12})(12) we get

\begin{eqnarray}\label{Eq.13}
d\Im&=&r_{0}^{2}\left(1+\sum_{\lambda\mu}a_{\lambda\mu}^{*}Y_{\lambda\mu}\left(\theta,\phi\right)\right)^{2}\frac{M}{4\pi}d\Omega \\ \nonumber
&=&r_{0}^{2}\left(1+2\sum_{\lambda\mu}a_{\lambda\mu}^{*}Y_{\lambda\mu}\left(\theta,\phi\right)\right. \\ \nonumber
&+&\left.\sum_{\lambda\lambda '\mu\mu '}a_{\lambda\mu}^{*}a_{\lambda '\mu '}^{*}Y_{\lambda\mu}Y_{\lambda '\mu '}\right)\frac{M}{4\pi}d\Omega.
\end{eqnarray}

Integrating both sides of ~Eq. (\ref{Eq.13})(13) over the whale surface of the nucleus, one can find

\begin{eqnarray}\label{Eq.14}
\Im&=&r_{0}^{2}\left(\int d\Omega+2\sum_{\lambda\mu}a_{\lambda\mu}^{*}\int Y_{\lambda\mu}\left(\theta,\phi\right)d\Omega\right. \\ \nonumber
&+&\left.\sum_{\lambda\lambda '\mu\mu '}a_{\lambda\mu}^{*}a_{\lambda '\mu '}^{*}\int Y_{\lambda\mu}Y_{\lambda '\mu '}d\Omega\right)\frac{M}{4\pi},
\end{eqnarray}

\noindent where $\int d\Omega=4\pi$, refer to the integration over solid angle. It is well known that the integration
$\int Y_{\lambda\mu}\left(\theta,\phi\right)d\Omega$ which is the second term in the above equation is zero and
$\int Y_{\lambda\mu}Y_{\lambda '\mu '}d\Omega=\left(-1\right)^{\mu}\delta_{\lambda\lambda '}\delta_{\mu,-\mu '}$.
The quantity $\left(-1\right)^{\mu}a_{\lambda '\mu '}^{*}$ equals to $a_{\lambda ',-\mu '}$. Putting all these
requirements into ~Eq. (\ref{Eq.14}) we get

\begin{eqnarray}\label{Eq.15}
\Im=\Im_{0}\left(1+\frac{1}{4\pi}\sum_{\lambda\mu}\left(-1\right)^{\mu}a_{\lambda,-\mu}a_{\lambda,\mu}\right),
\end{eqnarray}

\noindent
where $a_{\lambda\mu}$ is the deformation parameter in body-fixed coordinates. The second term in ~Eq. (\ref{Eq.15})
represents the departure from the spherical equilibrium shape in term of complete set of spherical harmonic
functions $Y_{\lambda\mu}$. It is found that even-even nuclei can be accurately described in terms of a
deformation of order $\lambda=2$. In ~Eq. (\ref{Eq.15}) the suffix index $\mu$ which represents the
orientation of the nucleus in space-fixed coordinates runs from $-\lambda$ to $\lambda$. In our case
$\mu$ runs $-2$ to $2$. We drop the subscript $\lambda=2$ from all deformation parameters henceforth

\begin{eqnarray}\label{Eq.16}
\Im=\Im_{0}\left(1+\frac{1}{4\pi}\sum_{\mu}\left(-1\right)^{\mu}a_{-\mu}a_{\mu}\right).
\end{eqnarray}

In the usual way $a_{\mu}$ can be written in term of the annihilation and creation operators of phonon
$\xi_{\mu}$ and $\xi_{\mu}^{T}$ as \cite{Landau013,Pal83, Greiner96}

\begin{eqnarray}\label{Eq.17}
a_{\mu}=\sqrt{\frac{\hbar\omega}{2C}}\left(\xi_{\mu}+(-1)^{\mu}\xi_{-\mu}^{T}\right),
\end{eqnarray}

\noindent
where $\omega=\sqrt{\frac{C}{B}}$ is the angular frequency, $B$, $C$ are the inertial and stiffness
parameters respectively. Putting ~Eq. (\ref{Eq.17}) into ~Eq. (\ref{Eq.16}) we get

\begin{eqnarray}\label{Eq.18}
\Im_{I}&=&\Im_{0}\left[1+\frac{1}{4\pi}\frac{\hbar\omega}{2C}\left(\sum_{\mu}\left(1+2\xi_{\mu}^{T}\xi_{\mu}\right)\right)\right] \\ \nonumber
&=&\Im_{0}\left[1+\frac{\hbar\omega}{8\pi C}\left(\sum_{\mu}\left(1+2\hat{n}_{\mu}\right)\right)\right],
\end{eqnarray}

\noindent
where $\hat{n}_{\mu}=\xi_{\mu}^{T}\xi_{\mu}$ is the number operator. Since ~Eq. (\ref{Eq.18})(17) include a parameter
represents the number operator it can be used not only for the ground state bands but also for beta bands and the first gamma bands.
In the case of the ground state band the number of operator is zero. Replacing $\hbar\omega$ by $\frac{\hbar^{2}I(I+1)}{2\Im_{I}}$
in ~Eq. (\ref{Eq.18}) is reduced to

\begin{eqnarray}\label{Eq.19}
\Im_{I}&=&\Im_{0}\left[1+5\frac{1}{8\pi C}\frac{\hbar^{2}I(I+1)}{2\Im_{I}}\right] \\ \nonumber
&=&\Im_{0}\left[1+\frac{5\hbar^{2}}{16\pi}\frac{I(I+1)}{\Im_{0}C\left[1+\frac{5\hbar^{2}}{16\pi}\frac{I(I+1)}{\Im_{I}C}\right]}\right] \\ \nonumber
&\cong&\Im_{0}\left[1+\frac{5\hbar^{2}}{16\pi}\frac{I(I+1)}{\Im_{0}C\left[1+\frac{5\hbar^{2}}{16\pi}\frac{I(I+1)}{\Im_{0}C}\right]}\right].
\end{eqnarray}

~Eq. (\ref{Eq.19}) is a recursion relation of $\Im_{I}$ (i.e., $\Im_{I}$ are defined in terms of itself), and it represents
the moment of inertia of the nucleus at an angular momentum $I$, approximated to the second order of perturbation.
The factor $5$ in ~Eq. (\ref{Eq.19}) arises because the summation over $\mu$ runs from $\lambda=-2$ to
$\lambda=2$ through $\mu=0$ as mentioned above. Finally, the energy levels in the ground state bands is casted
into the following form,

\begin{eqnarray}\label{Eq.20}
E_{rot}(I)&=&\frac{\hbar^{2}I(I+1)}{2\Im_{I}} \\ \nonumber
&=&\frac{\hbar^{2}I(I+1)}{2\Im_{0}\left[1+\frac{5\hbar^{2}}{16\pi}\frac{I(I+1)}{\Im_{0}C\left[1+\frac{5\hbar^{2}}{16\pi}\frac{I(I+1)}{\Im_{0}C}\right]}\right]} \\ \nonumber
&\equiv&\frac{A}{\left[1+\frac{BI(I+1)}{1+BI(I+1)}\right]}I(I+1),
\end{eqnarray}

\noindent where, $A\equiv\frac{\hbar^2}{2\Im_{0}}$ and $B\equiv\frac{5\hbar^2}{8\pi\Im_{0}C}$.
We will call this ~Eq. (\ref{Eq.20})(20) quantized $\beta$- stretching equation.

These parameters embed in them the intrinsic moment of inertia $\Im_{0}$ and stiffness of the nucleus $C$. They are constant for a
particular nucleus, but their values differ from one nucleus to another. In the present model, they are
adjustable parameters and evaluated by fitting ~Eq. (\ref{Eq.20})(20) against experimentally measured
energy levels. In the following section we discuss the validation of the model by comparing its predictions
against experimental data.



\section{\label{sec:level2} Results and discussion}

The present investigation is dedicated to study of the energy levels of the ground band states
in even–even isotopes $_{62}Sm$, $_{64}Gd$, $_{66}Dy$, $_{68}Er$, $_{70}Yb$, $_{72}Hf$, $_{74}W$ and $_{76}Os$
with the neutron numbers ranging from $90$ to $114$, by the quantizing of stretching model for 40 nuclei ranging
from atomic number $152$ to $190$ and having the energy ratio $2.9<R<3.33$

The simple expression ~Eq. (\ref{Eq.20}), has been used to evaluate the level energies up to spin $I=16$.
The parameters $A$ and $B$ in ~Eq. (\ref{Eq.20}) has been determined by the least squares method with fitting
first three energy values of ground state band, which is obtained experimentally ref. \cite{Http00}.

Comparison between the calculated results obtained by ~Eq. (\ref{Eq.20}) with the experimental data \cite{Http00}
taken from the decay data Webster (http://www.nndc.bnl.gov/nudat2/) and calculated values according to the $VMI$
model \cite{Mariscotti69} for the energy levels $E(I)$ of the ground state bands for a large collection of
even-even nuclei are given in Table 1.

It can be seen from Table 1 that there is an overall agreement with the experimental data with errors not more
than $0.5\%$ for most of the nuclei. Very few cases are found to display an error of $5\%$ which arises because
of the spread of the experimental points which increases rapidly with $I$. The present model has the advantage
of being simpler in the form of the energy levels than the $VMI$ model.

To understand the anomaly in ground state bands, we have carried out calculations and compared with the experimental
data ~\cite{Http00} which is taken from the decay data Webster (http://www.nndc.bnl.gov/nudat2/) and also with the
other theoretical results by the $VMI$ model ~\cite{Mariscotti69}, phenomenological model $Phen.M$ ~\cite{Okhunov015}
takes into account the Coriolis mixture of low-lying state bands for $^{170-176}Yb$, $^{170-180}Hf$, and $^{174-184}W$
nuclei, respectively as a detailed example, shown in Figure 1-3.


We can see from the Figure 1-3 that this comparison showed the results of our calculation are coincide with experimental
date and also with the results of $VMI$ and phenomenological $PhM$ models. And it can be seen that the calculated energy
levels reproduce the energies of the yrast levels qualitatively describe for all even-even nuclei.

A graphical comparison of the calculated and experimental values of the energy ratios $\frac{E_{I}}{E_{2}}$ of excited states
$\frac{E_{I}}{E_{2}}$ as a function of the ratio $R=\frac{E_{4}}{E_{2}}$ in the range $2.90$ to $3.33$ for all $I$ up to $16$
is shown in Figure 4.
It is clearly obvious that the coincidence between experimental data and the predictions of the current work is satisfactory.
Specifically, we observe that at energy levels where $I<14$ the theoretical curve represent with high accuracy the experimental
points. That is the theoretical curve passes nearly through all experimental points. For higher energy levels the spread of
the experimental points increases rapidly with $I$ and no longer a smooth curve to pass through all points.


Comparison of the ratio $E_{10}/E_{2}$ is plotted against $E_{4}/E_{2}$ which is obtained in the current model with the experimental
data \cite{Http00} and other three models, namely the Gupta [~Eq. (\ref{Eq.4})], Bohr-Mottelson 2-parameter models
[~Eq. (\ref{Eq.2})] and Sood [~Eq. (\ref{Eq.5})] are illustrated in Figure 5. While the prediction of our work
compares nicely with the experimental data for all the region under consideration of the nuclei, both Gupta and
Bohr-Mottelson 2 parameter model diverge from the experimental data except at the proximity around the
$\frac{E_{4}}{E_{2}}\approx3.33$ regime. This infers that the Gupta and Bohr-Mottelson 2 parameter model can
hold only a few nuclei which is confined in the range of $3.25<R<3.33$. Although the success of Sood is impressive
some shortcomings nonetheless. For example, a parameter $N$ is introduced [see ~Eq. (\ref{Eq.5})] without scientific
justification, nothing has been mentioned about the physical interpretation or identity of this parameter in spite
of its importance in fitting the experimental points.

\begin{widetext}
\begin{center}
\begin{table}[h!]
{\textbf{\caption {The energy levels $E(I)$ of the ground state band of even-even nuclei.
For each nucleus, the first row is the experimental values [http://www.nndc.bnl.gov/nudat2/], the second and third rows the calculated
values according to this work and that of the $VMI$ model \cite{Mariscotti69}.}}}
\begin{footnotesize}
\begin{tabular*} {0.90\textwidth}%
{@{\extracolsep{\fill}}cccccccccccc} \hline\hline
 Nuclei \ & \ A \ & \ B \ &  \ Results \ & \multicolumn{8}{c}{I}  \\ \cline{5-12}
 \ & \  & \ & \ & \ 2 \ & \ 4 \ & \ 6 \ & \ 8 \ & \ 10 \ & \ 12 \ & \ 14 \ & \ 16  \\ \hline
 \ $^{152}Sm$ \ & \ 21.76 \ & \ 0.021 \  & \ Exp. \ & \ 121.8 \ & \ 366.5 \ & \ 706.9 \ & \ 1125.4 \ & \ 1609.3 \ & \ 2148.8 \ & \ 2736.2 \ & \ 3365.0  \\
  & & & \ $VMI$ \ & \ 121.0 \ & \ 369.9 \ & \ 712.3 \ & \ 1127.3 \ & \ 1601.8 \ & \ 2127.2 \ & \ 2697.2 \ & \ 3307.0  \\
  & & & \ Current work \ & \ 114.0 \ & \ 354.9 \ & \ 698.1 \ & \ 1121.4 \ & \ 1614.3 \ & \ 2157.3 \ & \ 2741.2 \ & \ 3356.9  \\ \hline

\ $^{154}Sm$ \ & \ 14.47 \ & \ 0.0088  \ & \ Exp.  \ & \ 82.0 \ & \ 266.8 \ & \ 544.1 \ & \ 902.8 \ & \ 1333.0 \ & \ 1825.9 \ & \ 2373.0 \ & \ 2968.2  \\
  & & & \ $VMI$ \              & \ 81.5 \ & \ 267.7 \ & \ 550.4 \ & \ 920.3 \ & \ 1368.3 \ & \ 1886.5 \ & \ 2468.2 \ & \ 3207.7  \\
  & & & \ Current work \       & \ 82.8 \ & \ 267.1 \ & \ 543.4 \ & \ 903.0 \ & \ 1338.0 \ & \ 1841.0 \ & \ 2406.0 \ & \ 3026.0  \\ \hline\hline

\ $^{154}Gd$ \ & \  21.99 \ & \ 0.0206  \ & \ Exp.  \ & \ 123.1 \ & \ 371.0 \ & \ 717.7 \ & \ 1144.4 \ & \ 1637.1 \ & \ 2184.7 \ & \ 2777.3 \ & \ 3404.5  \\
  & & & \ $VMI$ \              & \ 122.0 \ & \ 374.4 \ & \ 722.8 \ & \ 1146.0 \ & \ 1630.7 \ & \ 2167.8 \ & \ 2750.9 \ & \ 3375.1  \\
  & & & \ Current work \       & \ 120.7 \ & \ 370.6 \ & \ 719.4 \ & \ 1143.7 \ & \ 1624.9 \ & \ 2148.5 \ & \ 2770.3 \ & \ 3279.3  \\ \hline

\ $^{156}Gd$ \ & \ 15.87 \ & \ 0.0105  \ & \ Exp.  \ & \ 89.0 \ & \ 288.2 \ & \ 584.7 \ & \ 965.1 \ & \ 1416.1 \ & \ 1924.5 \ & \ 2475.8 \ & \ 3059.5  \\
  & & & \ $VMI$ \              & \ 88.8 \ & \ 288.4 \ & \ 585.0 \ & \ 965.2 \ & \ 1417.9 \ & \ 1934.2 \ & \ 2506.9 \ & \ 3130.7  \\
  & & & \ Current work \       & \ 89.8 \ & \ 288.1 \ & \ 582.7 \ & \ 962.9 \ & \ 1419.0 \ & \ 1942.0 \ & \ 2525.5 \ & \ 3162.3  \\ \hline

\ $^{158}Gd$ \ & \ 13.82 \ & \ 0.0058  \ & \ Exp.  \ & \ 79.5 \ & \ 261.5 \ & \ 539.0 \ & \ 904.1 \ & \ 1349.0 \ & \ 1865.0 \ & \ -- \ & \ --  \\
  & & & \ $VMI$ \              & \ 79.6 \ & \ 261.4 \ & \ 537.8 \ & \ 899.5 \ & \ 1338.0 \ & \ 1845.4 \ & \ 2415.3 \ & \ 3042.1  \\
  & & & \ Current work \       & \ 80.1 \ & \ 261.4 \ & \ 537.8 \ & \ 903.2 \ & \ 1352.2 \ & \ 1879.7 \ & \ 2480.6 \ & \ 3150.5  \\ \hline

\ $^{160}Gd$ \ & \ 13.00 \ & \ 0.0046  \ & \ Exp. \ & \ 75.3 \ & \ 248.5 \ & \ 514.8 \ & \ 867.9 \ & \ 1300.7 \ & \ 1806.3 \ & \ 2377.3 \ & \ 3007.1  \\
  & & & \ $VMI$ \              & \ 75.8 \ & \ 247.5 \ & \ 511.5 \ & \ 860.0 \ & \ 1285.7 \ & \ 1781.8 \ & \ 2342.2 \ & \ 2961.7  \\
  & & & \ Current work \       & \ 75.8 \ & \ 248.8 \ & \ 514.3 \ & \ 868.1 \ & \ 1306.0 \ & \ 1824.1 \ & \ 2418.6 \ & \ 3058.9   \\ \hline\hline

\ $^{156}Dy$ \ & \ 24.60 \ & \ 0.0240  \ & \ Exp. \ & \ 334.3 \ & \ 746.8 \ & \ 1223.7 \ & \ 1747.2 \ & \ 2304.1 \ & \ 2892.6 \ & \ 3508.7 \ & \ 4172.8  \\
  & & & \ $VMI$        \ & \ 333.7 \ & \ 744.8 \ & \ 1221.5 \ & \ 1749.7 \ & \ 2321.1 \ & \ 2929.9 \ & \ 3572.1 \ & \ 4244.6   \\
  & & & \ Current work \ & \ 297.0 \ & \ 754.3 \ & \ 1248.4 \ & \ 1731.7 \ & \ 2184.0 \ & \ 2601.2 \ & \ 2982.0 \ & \ 3328.8   \\ \hline

\ $^{158}Dy$ \ & \ 17.81 \ & \ 0.0129  \ & \ Exp. \ & \ 79.5 \ & \ 261.5 \ & \ 539.0 \ & \ 904.1 \ & \ 1349.0 \ & \ 1865.0 \ & \ -- \ & \ --  \\
  & & & \ $VMI$        \ & \ 79.6 \ & \ 261.4 \ & \ 537.8 \ & \ 899.5 \ & \ 1338.0 \ & \ 1845.4 \ & \ 2415.3 \ & \ 3042.1   \\
  & & & \ Current work \ & \ 80.1 \ & \ 261.4 \ & \ 537.8 \ & \ 903.2 \ & \ 1352.2 \ & \ 1879.7 \ & \ 2480.6 \ & \ 3150.5   \\ \hline

\ $^{160}Dy$ \ & \ 15.37 \ & \ 0.0084 \ & \ Exp. \  & \ 86.8 \ & \ 283.8 \ & \ 581.1 \ & \ 966.9 \ & \ 1428.0 \ & \ 1950.5 \ & \ 2513.8 \ & \ 3089.8  \\
  & & & \ $VMI$        \ & \ 86.7 \ & \ 284.0 \ & \ 582.6 \ & \ 971.7 \ & \ 1441.5 \ & \ 1983.4 \ & \ 2590.2 \ & \ 3256.0   \\
  & & & \ Current work \ & \ 88.6 \ & \ 285.6 \ & \ 580.6 \ & \ 964.8 \ & \ 1429.0 \ & \ 1965.6 \ & \ 2567.7 \ & \ 3228.6   \\ \hline

\ $^{162}Dy$ \ & \ 14.02 \ & \ 0.0056 \ & \ Exp. \  & \ 80.7 \ & \ 265.7 \ & \ 548.5 \ & \ 921.3 \ & \ 1375.1 \ & \ 1901.1 \ & \ 2491.7 \ & \ 3138.6  \\
  & & & \ $VMI$        \ & \ 80.9 \ & \ 266.2 \ & \ 549.2 \ & \ 921.5 \ & \ 1374.8 \ & \ 1901.6 \ & \ 2495.3 \ & \ 3150.2   \\
  & & & \ Current work \ & \ 81.9 \ & \ 266.9 \ & \ 548.4 \ & \ 920.0 \ & \ 1375.7 \ & \ 1910.2 \ & \ 2518.3 \ & \ 3195.0   \\ \hline

\ $^{164}Dy$ \ & \ 12.74 \ & \ 0.0051 \ & \ Exp. \  & \ 73.4 \ & \ 242.2 \ & \ 501.3 \ & \ 843.7 \ & \ 1261.3 \ & \ 1745.9 \ & \ 2290.6 \ & \ 2887.1  \\
  & & & \ $VMI$        \ & \ 73.5 \ & \ 242.1 \ & \ 500.1 \ & \ 840.3 \ & \ 1255.4 \ & \ 1738.7 \ & \ 2284.2 \ & \ 2886.8   \\
  & & & \ Current work \ & \ 74.6 \ & \ 243.5 \ & \ 501.2 \ & \ 842.3 \ & \ 1262.0 \ & \ 1755.4 \ & \ 2318.2 \ & \ 2946.1   \\ \hline\hline

\ $^{162}Er$ \ & \ 18.32 \ & \ 0.0115 \ & \ Exp. \  & \ 102.0 \ & \ 329.6 \ & \ 666.7 \ & \ 1096.7 \ & \ 1602.8 \ & \ 2165.1 \ & \ 2745.7 \ & \ 3292.4  \\
  & & & \ $VMI$        \ & \ 101.0 \ & \ 327.1 \ & \ 661.7 \ & \ 1089.1 \ & \ 1596.4 \ & \ 2173.5 \ & \ 2812.7 \ & \ 3507.7   \\
  & & & \ Current work \ & \ 103.9 \ & \ 331.3 \ & \ 666.1 \ & \ 1094.3 \ & \ 1603.7 \ & \ 2183.5 \ & \ 2824.3 \ & \ 3517.7   \\ \hline

\ $^{164}Er$ \ & \ 16.09 \ & \ 0.0076 \ & \ Exp. \  & \ 91.4 \ & \ 299.4 \ & \ 614.4 \ & \ 1024.6 \ & \ 1518.1 \ & \ 2082.8 \ & \ 2702.6 \ & \ 3411.2  \\
  & & & \ $VMI$        \ & \ 90.9 \ & \ 297.6 \ & \ 610.0 \ & \ 1016.7 \ & \ 1507.7 \ & \ 2072.3 \ & \ 2704.7 \ & \ 3398.2   \\
  & & & \ Current work \ & \ 93.1 \ & \ 301.1 \ & \ 614.1 \ & \ 1022.8 \ & \ 1519.0 \ & \ 2094.9 \ & \ 2743.4 \ & \ 3458.1   \\ \hline

\ $^{166}Er$ \ & \ 14.25 \ & \ 0.0078 \ & \ Exp. \  & \ 80.6 \ & \ 265.0 \ & \ 545.5 \ & \ 911.2 \ & \ 1349.5 \ & \ 1846.5 \ & \ 2389.3 \ & \ 2967.3  \\
  & & & \ $VMI$        \ & \ 80.6 \ & \ 264.8 \ & \ 544.6 \ & \ 910.6 \ & \ 1354.2 \ & \ 1867.3 \ & \ 2443.5 \ & \ 3077.0   \\
  & & & \ Current work \ & \ 82.5 \ & \ 267.2 \ & \ 545.2 \ & \ 908.8 \ & \ 1350.7 \ & \ 1864.0 \ & \ 2442.7 \ & \ 3081.0   \\ \hline

\ $^{168}Er$ \ & \ 13.68 \ & \ 0.0036 \ & \ Exp. \  & \ 79.8 \ & \ 264.1 \ & \ 548.7 \ & \ 928.3 \ & \ 1396.8 \ & \ 1947.3 \ & \ 2571.3 \ & \ 3259.5  \\
  & & & \ $VMI$        \ & \ 79.8 \ & \ 264.1 \ & \ 548.9 \ & \ 928.9 \ & \ 1398.0 \ & \ 1950.2 \ & \ 2579.6 \ & \ 3281.1   \\
  & & & \ Current work \ & \ 80.6 \ & \ 265.0 \ & \ 548.6 \ & \ 927.4 \ & \ 1397.2 \ & \ 1954.3 \ & \ 2595.1 \ & \ 3315.8   \\ \hline

\ $^{170}Er$ \ & \ 13.47 \ & \ 0.0036 \ & \ Exp. \  & \ 78.6 \ & \ 260.1 \ & \ 540.7 \ & \ 915.0 \ & \ 1376.6 \ & \ 1918.6 \ & \ 2537.2 \ & \ 3225.7  \\
  & & & \ $VMI$        \ & \ 79.0 \ & \ 261.1 \ & \ 541.9 \ & \ 915.3 \ & \ 1374.8 \ & \ 1914.0 \ & \ 2526.9 \ & \ 3208.0   \\
  & & & \ Current work \ & \ 79.5 \ & \ 261.1 \ & \ 540.6 \ & \ 913.9 \ & \ 1377.1 \ & \ 1926.3 \ & \ 2558.0 \ & \ 3268.7   \\ \hline\hline

\ $^{182}Os$ \ & \ 24.04 \ & \ 0.0197  \ & \ Exp. \ & \ 126.9 \ & \ 400.3 \ & \ 794.0 \ & \ 1277.9 \ & \ 1812.0 \ & \ 2346.1 \ & \ 2840.7 \ & \ 3320.1  \\
  & & & \ $VMI$        \ & \ 127.3 \ & \ 400.5 \ & \ 788.7 \ & \ 1268.9 \ & \ 1825.8 \ & \ 2448.5 \ & \ 3129.0 \ & \ 3861.2   \\
  & & & \ Current work \ & \ 131.2 \ & \ 406.0 \ & \ 793.4 \ & \ 1270.3 \ & \ 1815.8 \ & \ 2414.7 \ & \ 3054.2 \ & \ 3724.0   \\ \hline

\ $^{184}Os$ \ & \ 21.17 \ & \ 0.0110  \ & \ Exp. \ & \ 119.8 \ & \ 383.7 \ & \ 774.1 \ & \ 1274.8 \ & \ 1871.2 \ & \ 2547.6 \ & \ 3261.4 \ & \ 4046.5  \\
  & & & \ $VMI$        \ & \ 119.4 \ & \ 385.0 \ & \ 775.4 \ & \ 1271.1 \ & \ 1856.8 \ & \ 2520.8 \ & \ 3254.2 \ & \ 4049.8   \\
  & & & \ Current work \ & \ 120.2 \ & \ 384.1 \ & \ 774.0 \ & \ 1274.4 \ & \ 1871.4 \ & \ 2553.1 \ & \ 3308.7 \ & \ 4128.6   \\ \hline

\ $^{186}Os$ \ & \ 24.41 \ & \ 0.0133  \ & \ Exp. \ & \ 137.2 \ & \ 434.1 \ & \ 868.9 \ & \ 1420.9 \ & \ 2068.0 \ & \ 2781.3 \ & \ 3557.7 \ & \ --  \\
  & & & \ $VMI$        \ & \ 136.6 \ & \ 436.3 \ & \ 870.6 \ & \ 1415.8 \ & \ 2054.5 \ & \ 2774.0 \ & \ 3564.8 \ & \ 4419.5   \\
  & & & \ Current work \ & \ 137.4 \ & \ 435.0 \ & \ 869.2 \ & \ 1419.6 \ & \ 2068.6 \ & \ 2801.3 \ & \ 3604.6 \ & \ 4467.3   \\ \hline

\ $^{188}Os$ \ & \ 28.04 \ & \ 0.0183  \ & \ Exp. \ & \ 155.0 \ & \ 478.0 \ & \ 940.0 \ & \ 1514.8 \ & \ 2170.1 \ & \ 2856.3 \ & \ 3562.6 \ & \ 4236.5  \\
  & & & \ $VMI$        \ & \ 154.1 \ & \ 481.4 \ & \ 941.5 \ & \ 1506.9 \ & \ 2159.6 \ & \ 2886.9 \ & \ 3679.8 \ & \ 4531.3   \\
  & & & \ Current work \ & \ 154.3 \ & \ 479.1 \ & \ 941.1 \ & \ 1512.6 \ & \ 2171.0 \ & \ 2898.0 \ & \ 3678.5 \ & \ 4500.3   \\ \hline

\ $^{190}Os$ \ & \ 33.44 \ & \ 0.0238  \ & \ Exp. \ & \ 186.7 \ & \ 547.9 \ & \ 1050.4 \ & \ 1666.5 \ & \ 2357.0 \ & \ -- \ & \ -- \ & \ --  \\
  & & & \ $VMI$        \ & \ 185.3 \ & \ 554.6 \ & \ 1052.4 \ & \ 1648.5 \ & \ 2325.0 \ & \ 3069.8 \ & \ 3874.8 \ & \ 4733.4   \\
  & & & \ Current work \ & \ 179.1 \ & \ 546.3 \ & \ 1054.3 \ & \ 1666.7 \ & \ 2356.0 \ & \ 3100.4 \ & \ 3883.4 \ & \ 4691.9   \\ \hline\hline

\end{tabular*}
\end{footnotesize}
\end{table}
\end{center}

\begin{figure}[htp] \label{Fig.1}
\includegraphics[width=0.8\textwidth]{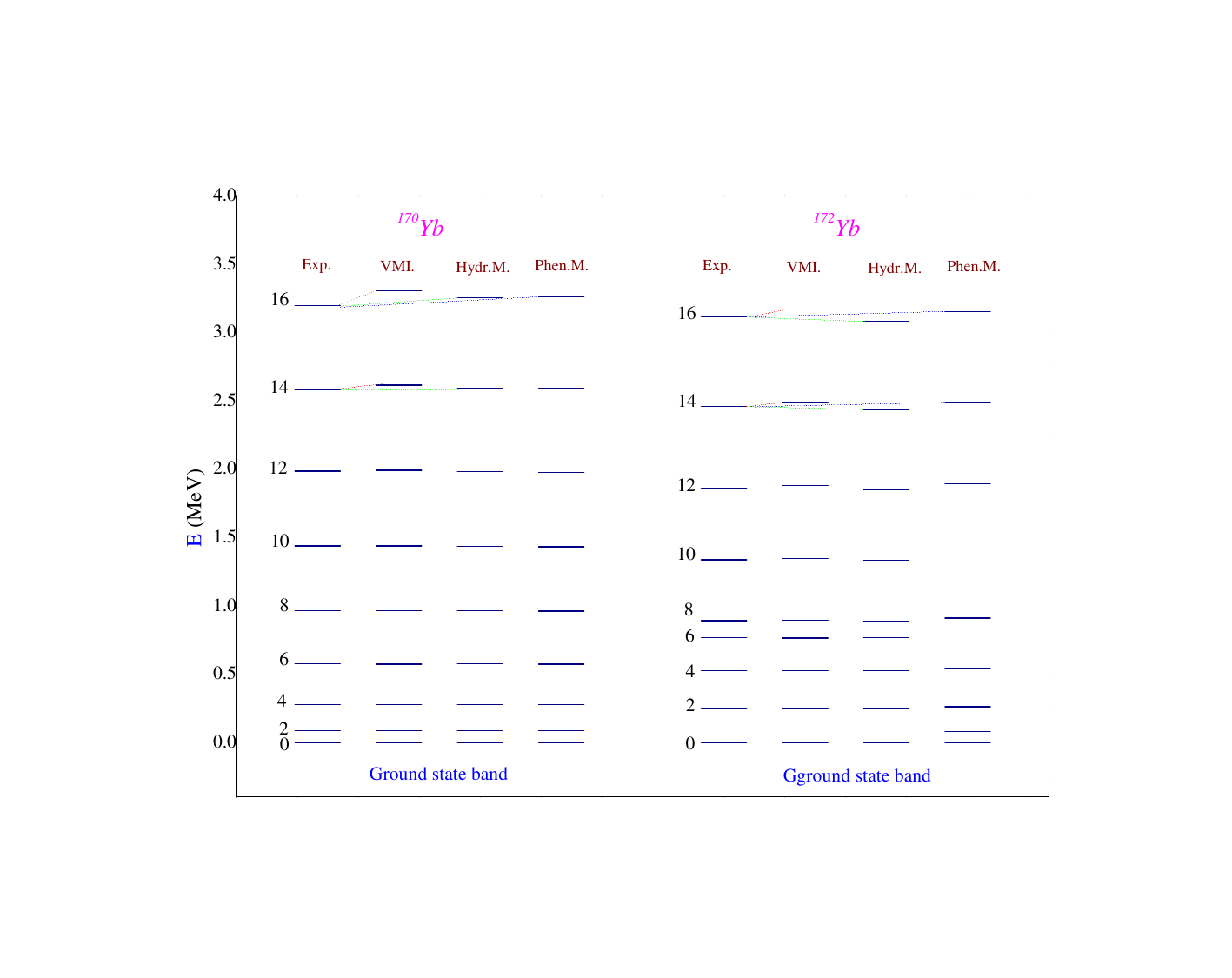} \\ \includegraphics[width=0.8\textwidth]{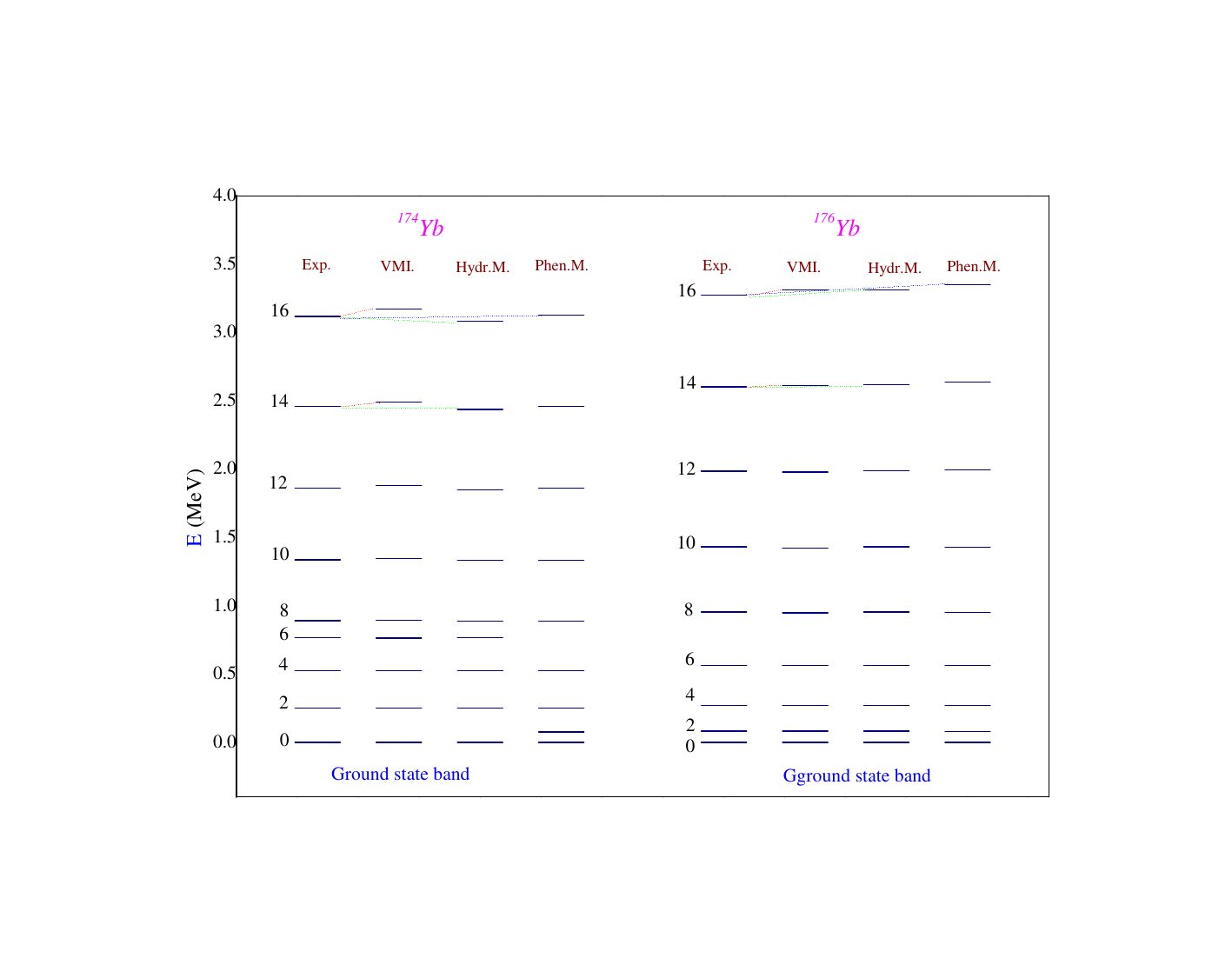}
  \caption{(Color online) Comparison between calculated results and experimental data ~\cite{Http00}, moreover with the theoretical results obtained by the $VMI$ ~\cite{Mariscotti69}, and $Phen.M$ ~\cite{Okhunov015} of the energy spectra of ground state bands in isotopes $Yb$.}
\end{figure}

\begin{figure}[htp] \label{Fig.2}
\includegraphics[width=0.8\textwidth]{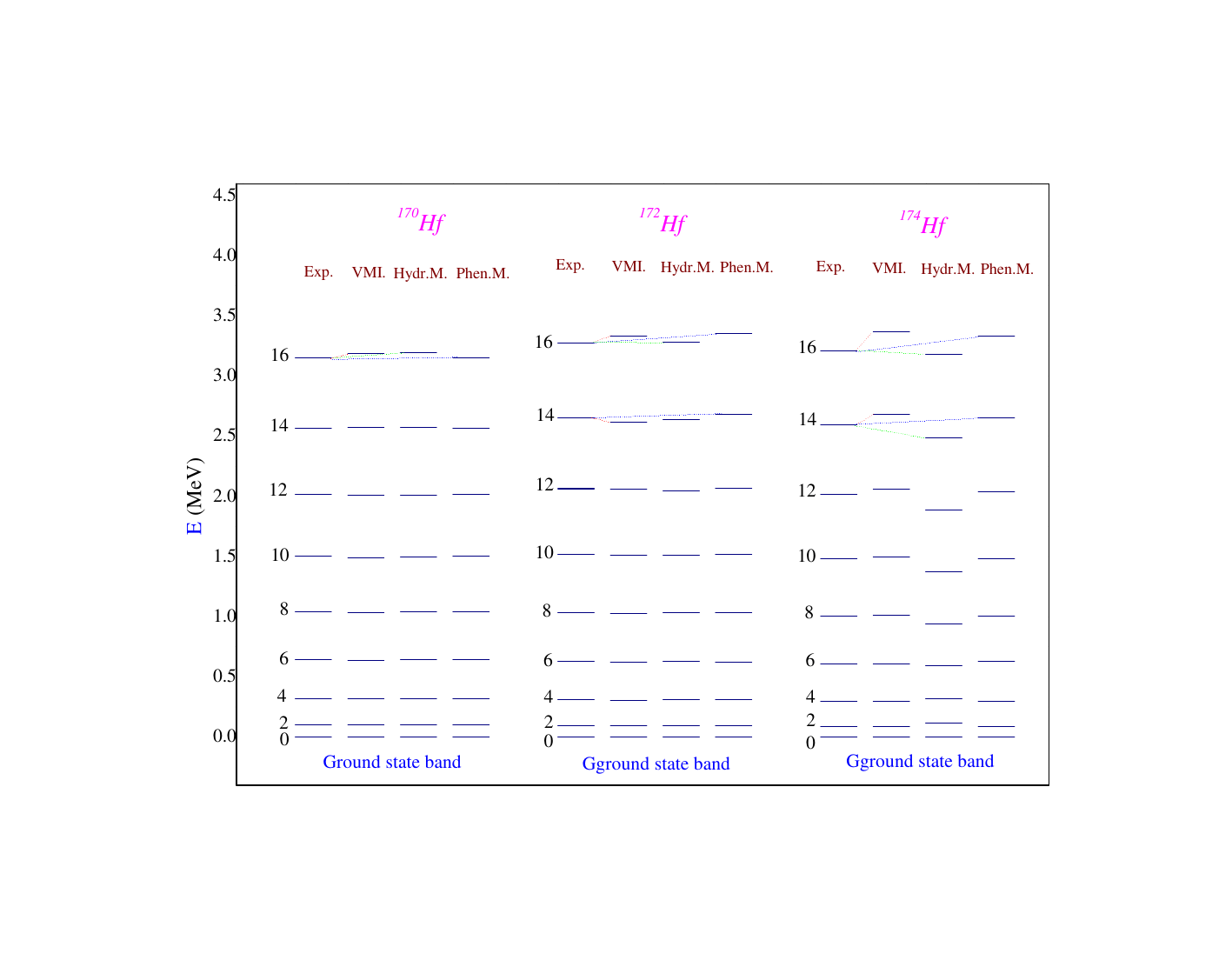} \\ \includegraphics[width=0.8\textwidth]{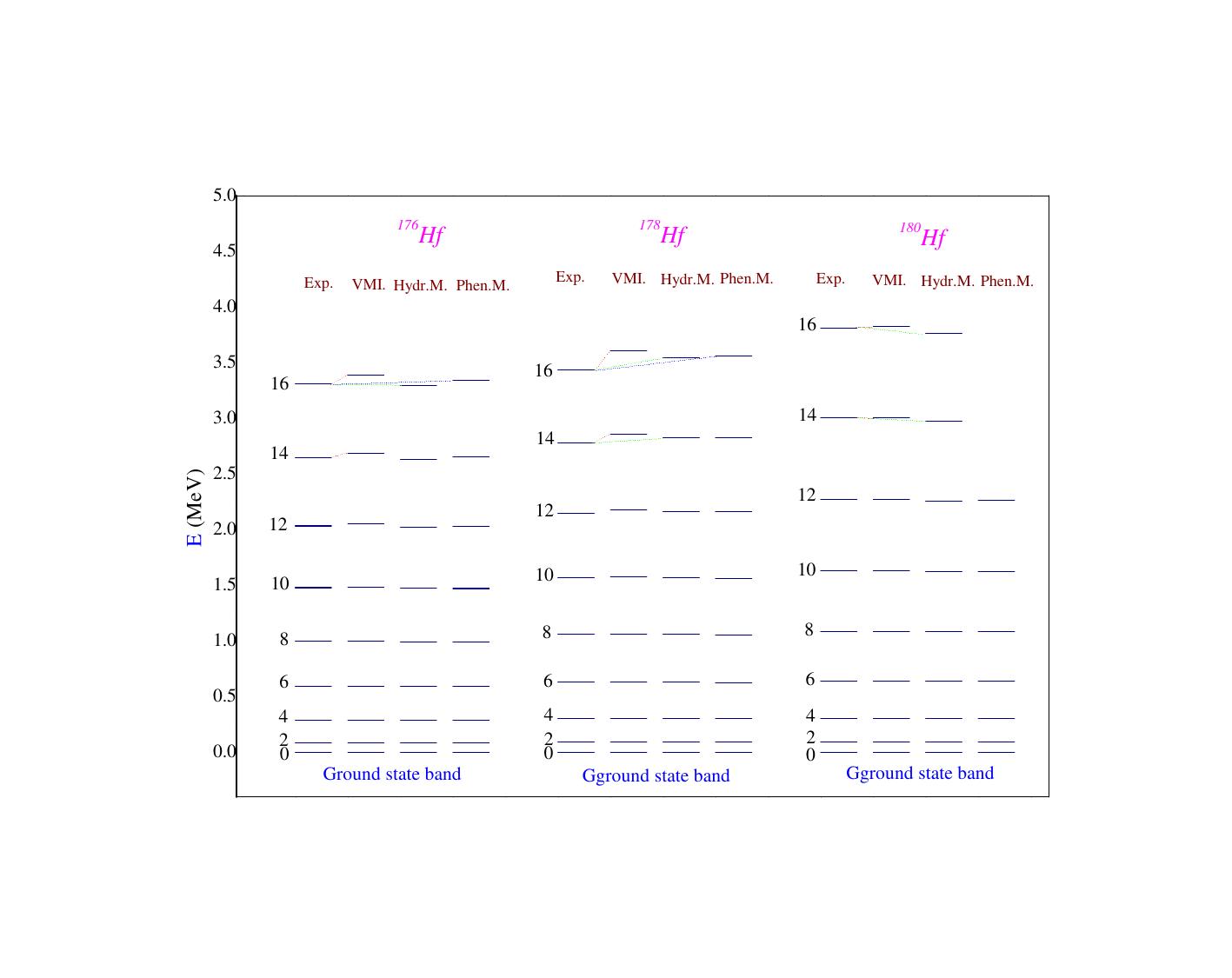}
  \caption{(Color online) Comparison between calculated results and experimental data ~\cite{Http00}, moreover with the theoretical results obtained by the $VMI$ ~\cite{Mariscotti69}, and $Phen.M$ ~\cite{Okhunov015} of the energy spectra of ground state bands in isotopes $Hf$.}
\end{figure}

\begin{figure}[htp] \label{Fig.3}
\includegraphics[width=0.8\textwidth]{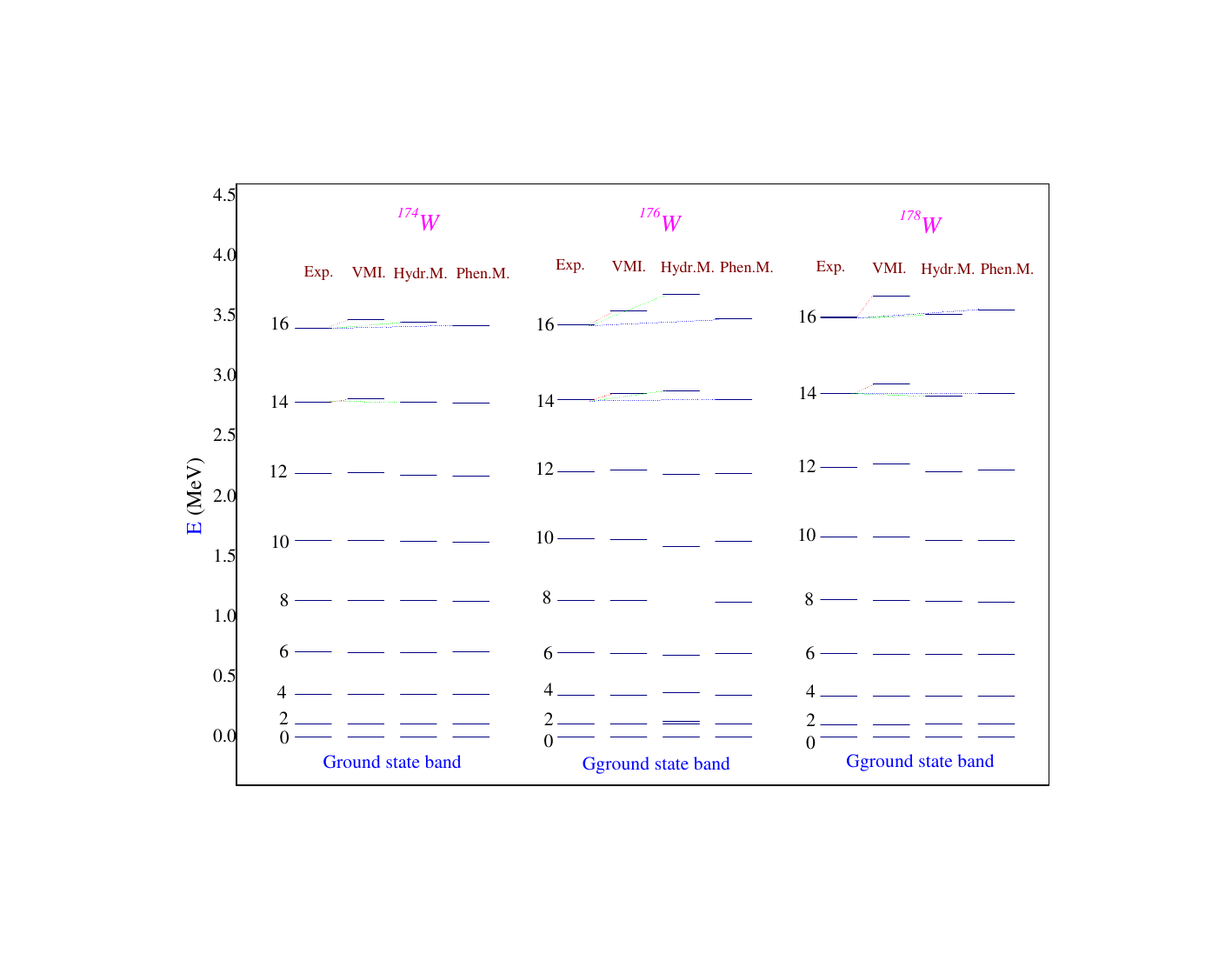} \\ \includegraphics[width=0.8\textwidth]{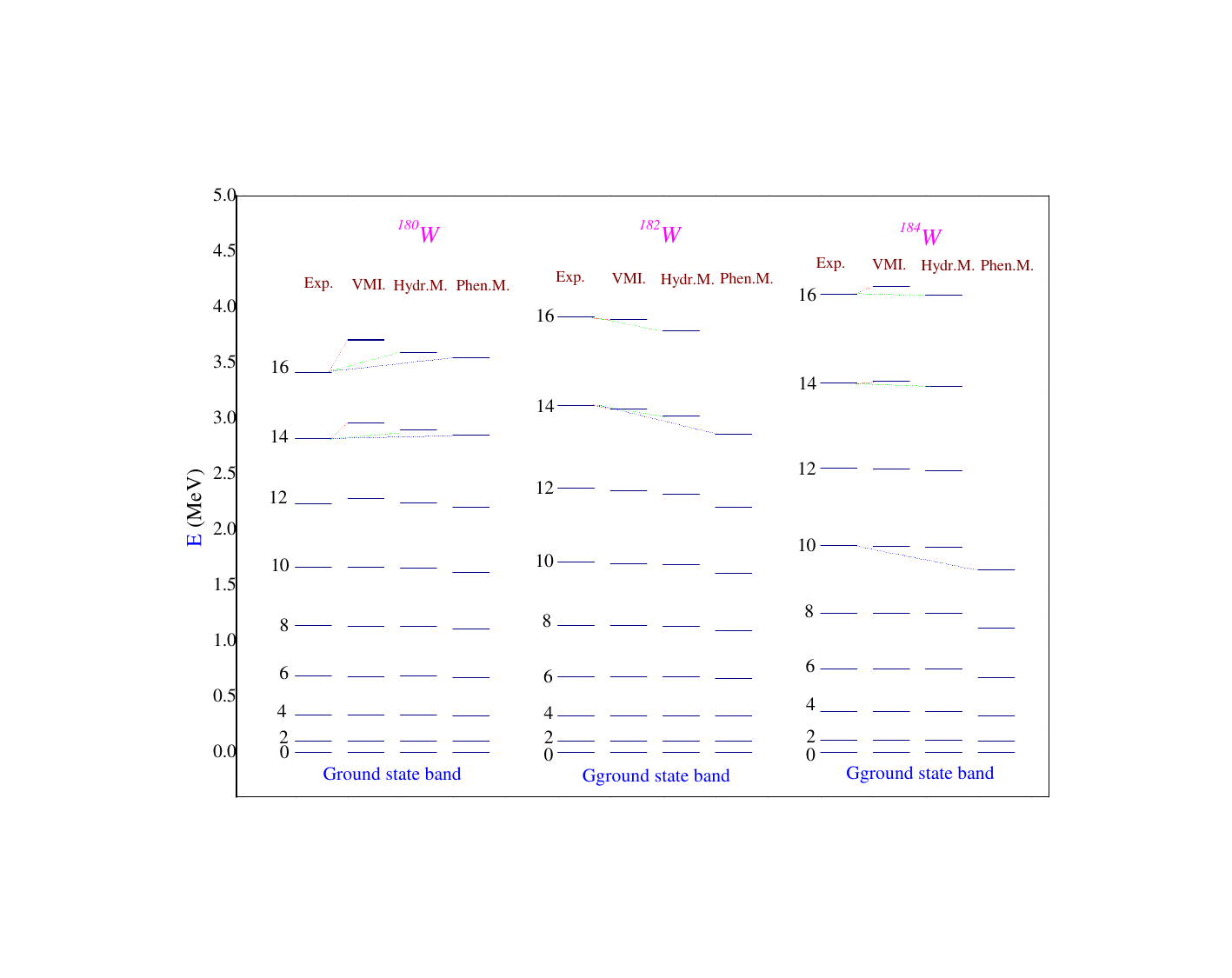}
  \caption{(Color online) Comparison between calculated results and experimental data ~\cite{Http00}, moreover with the theoretical results obtained by the $VMI$ ~\cite{Mariscotti69}, and $Phen.M$ ~\cite{Okhunov015} of the energy spectra of ground state bands in isotopes $W$.}
\end{figure}

\begin{figure}[htp] \label{Fig.4}
  \includegraphics[width=0.8\textwidth]{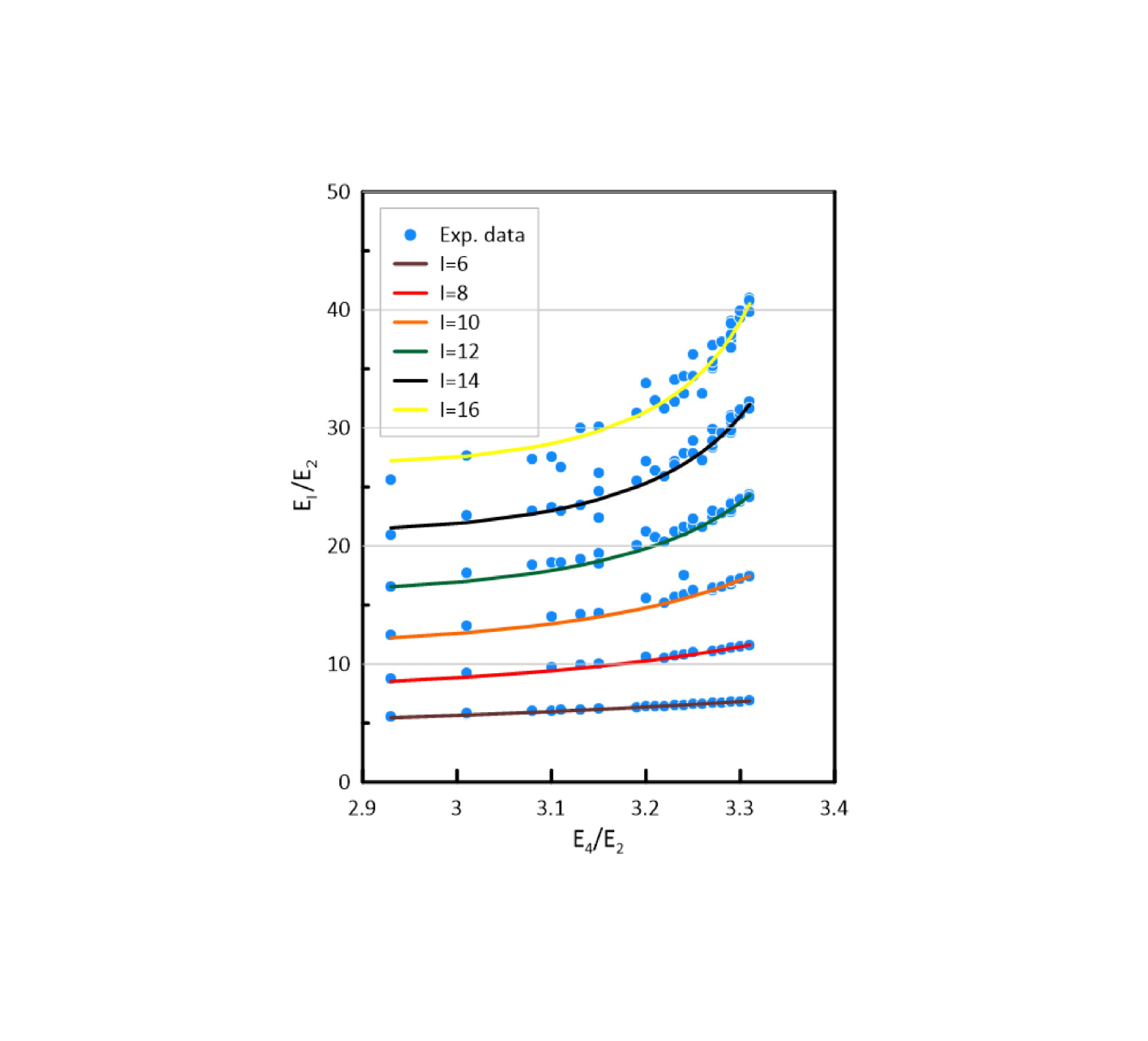}
  \caption{(Color online) Comparison calculated results obtained by ~Eq. (\ref{Eq.15}) with the experimental data \cite{Mariscotti69} for
energy ratio $E_{I}/E_{2}$ as a function of $R$ for a different values of $I$.}
\end{figure}

\begin{figure}[htp] \label{Fig.5}
  \includegraphics[width=0.8\textwidth]{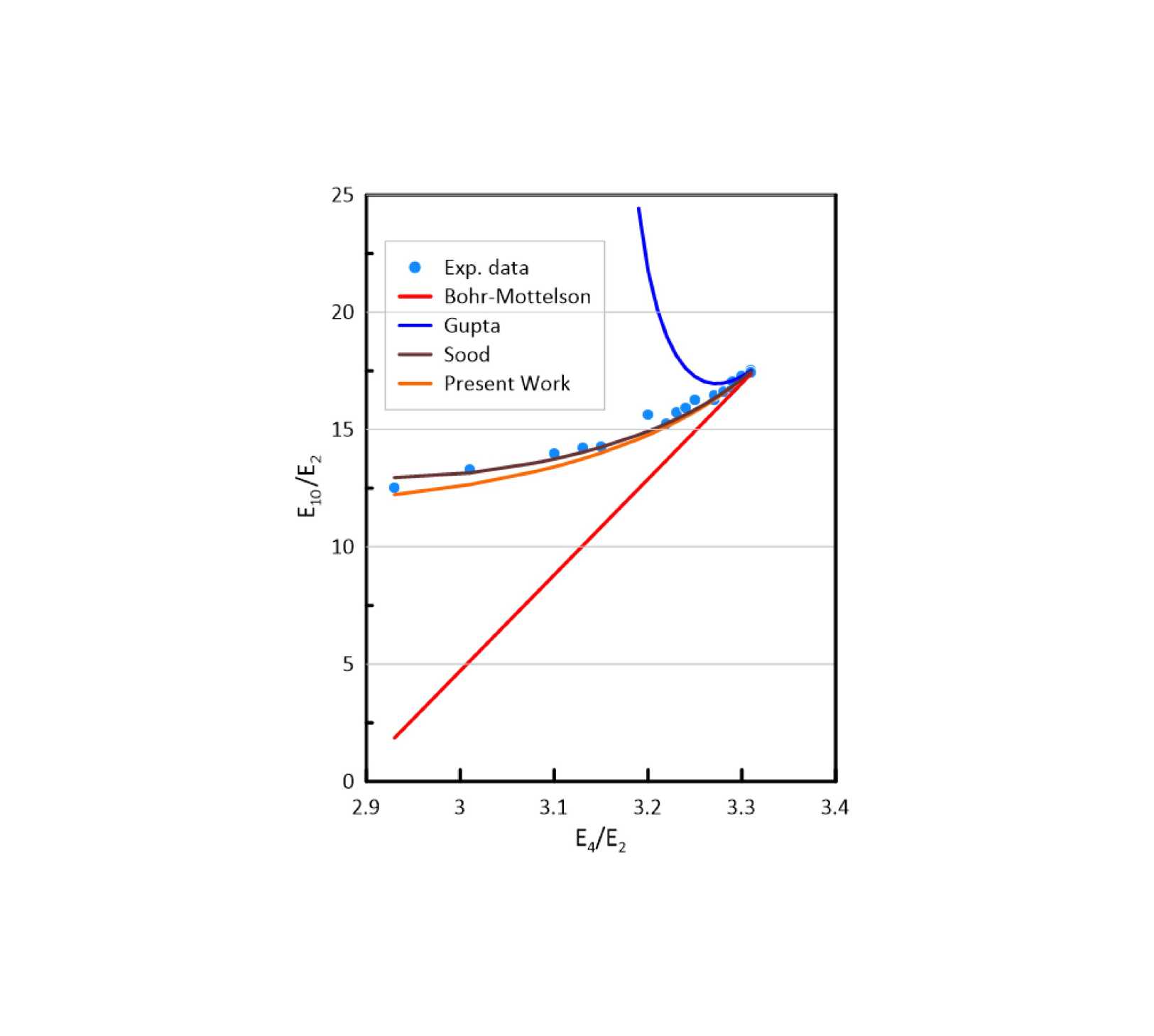}
  \caption{(Color online) Comparison of calculated results with the experimental data and results of other
models for the energy ratio $\frac{E_{10}}{E_{2}}$ as a function of $R=\frac{E_{4}}{E_{2}}$.
The solid curves are computed according ~Eq. (15) and dot is experimental data \cite{Mariscotti69}.}
\end{figure}
\end{widetext}

\section{\label{sec:level2} Summary and Conclusion}

The energy spectra of the ground state bands of even–even $_{62}Sm$, $_{64}Gd$, $_{66}Dy$, $_{68}Er$, $_{70}Yb$,
$_{72}Hf$, $_{74}W$ and $_{76}Os$ isotopes have been systematically studied using the theoretical framework of the
stretching model where the stretching of the nucleus with the rotation is quantized. A semi-classical formula for
the ground states bands of even-even nuclei is obtained. This derivation is made within the framework of the stretching
model where the stretching of the nucleus with the rotation is quantized.

\begin{itemize}
\item The quantization of the stretching that spinning nuclei exhibit led to a formula for the moment of inertia that
can be applied successfully for all nuclei $\frac{E_4}{E_2}\geq2.9$. While the classical treatment of such stretching
in References ~\cite{Diamond64} works well only for nuclei in the region of strongly deformed nuclei where $\frac{E_4}{E_2}\geq3.2$.
\item Unlike $VMI$ where the moment of inertia $\Im$ was considered as a general variable, our work has a characteristic that
the two parameters in our formula are clear and their physical meanings are well known.
\item The full coincidence of our formula which is derived on the basis that the moment of inertia depends only on the
deformation parameter with the results presented in references ~\cite{Okhunov015,Mariscotti69} which have taken into
consideration, in addition to deformation parameter, the pairing effect and Coriolis interaction means that the
contributions of the latter two parameters, i.e. pairing effect and Coriolis interaction are so small and it can be neglected.
\item The excellent fitting that this formula with the experimental results especially in levels with $I\leq14$ insure that the low-lying energy levels represent a collective motion of the nuclear surface. For high energy levels there are small deviations of the experimental value from the theoretical curve due to small contribution from microscopic features. This means that microscopic considerations should be taken into account in order to get a very accuracy results at high energy levels.
\end{itemize}

\begin{acknowledgments}
This work has been financial supported by the MOHE, Fundamental
Research Grant Scheme FRGS19-039-0647 and also OT$\sim$F2$\sim$2017/2020
by the Committee for the Coordination of the Development of Science and
Technology under the Cabinet of Ministers of the Republic of Uzbekistan.
We thank to Prof. Dr. P.N. Usmanov for discussing what helped us to improve our calculations
\end{acknowledgments}

\newpage

\begin{thebibliography}{1000}

\bibitem{Bohr51} Bohr A. {\it Nuclear magnetic moments and atomic hyperfine structure}. Physical Review 1951, {\bf 81}(3): 331.

\bibitem{Bohr52} A.Bohr "{\it The coupling of nuclear surface oscillations to
the motion of individual nucleons.}, Munksgaard: (1952).

\bibitem{Bohr76} A.Bohr {\it Rotational motion in nuclei.}
// Reviews of Modern Physics {\bf 48}, 365 (1976).

\bibitem{Bohr53} A.Bohr and B.R.Mottelson {\it Rotational states in even-even nuclei.}
// Physical Review {\bf 90}(4), 717 (1953).

\bibitem{Scharff55} G.Scharff-Goldhaber, J.Weneser {\it System of even-even nuclei.}
// Physical Review {\bf 98}, 212 (1955).

\bibitem{Wilets56} L.Wilets, M.Jean {\it Surface oscilations in even-even nuclei.}
// Physical Review {\bf 102}, 788 (1956).

\bibitem{Sorensen73} R.A.Sorensen {\it Nuclear momen of inertia at hifh spin.}
// Reviews of ModernPhysics {\bf 45}, 533 (1973).

\bibitem{Preston018} M.A.Preston {\it Structure of the Nucleus.}
// CRC Press: (2018).

\bibitem{Sood67} P.Sood, {\it Semiempirical formula for nuclear rotational energies.}
// Physical Review {\bf 161}(4), 1063 (1967).

\bibitem{Gupta67} R.K.Gupta {\it Rotational States in Deformed Even–Even Nuclei.}
// Canadian Journal of Physics {\bf 45}(11), 3521-3532 (1967).

\bibitem{Gupta69} R.K.Gupta {\it Higher order corrections to the rigid rotator law I(I+1).}
// Canadian Journal of Physics {\bf 47}(3), 299-307 (1969).

\bibitem{Trainor71} I.Trainor, R.K.Gupta {\it Rotational Invariance in the Centrifugal Stretching of Deformed Nuclei.}
// Canadian Journal of Physics {\bf 49}(1), 133-143 (1971).

\bibitem{Sood68} P.Sood {\it Centrifugal stretching of a classical rotator and collective motions in nuclei.}
// Canadian Journal of Physics {\bf 46}(12), 1419-1423 (1968).

\bibitem{Volkov71} A.Volkov {\it A note on the analysis of rotational spectra.}
// Physics Letter B {\bf 35}(4), 299-302 (1971).

\bibitem{Morinaga66} H.Morinaga {\it Rotational bands of well-deformed nuclei studied from gamma rays following ($\alpha$, $xn$) reactions.}
// Nuclear Physics {\bf 75}, 385 (1966).

\bibitem{Diamond64} R.Diamond, F.Stephens, W.Swiatecki {\it Centrifugal stretching of nuclei.} (1964).

\bibitem{Mariscotti69} M.Mariscotti, G.Scharff-Goldhaber, B.Buck
{\it Phenomenological analysis of ground-state bands in even-even nuclei.}
// Physical Review {\bf 178}(4), 1864 (1969).

\bibitem{Scharff76} G. Scharff-Goldhaber, C. B. Dover, A. L. Goodman {\it The Variable Moment of Inertia ($VMI$) Model
and Theories of Nuclear Collective Motion.} // Annual reviwew of nuclear science {\bf 26}, 239 (1976).

\bibitem{Harris65} S. M. Harris {\it Higher order correction to the Cranking Model}, Phys. Rev. B, {\bf 138}, 509--513 (1965).

\bibitem{Okhunov015} A. A.Okhunov, G. I.Turaeva, H. A.Kassim, M. U. Khandaker, N. B. Rosli {\it Analysis of the energy spectra of
ground states of deformed nuclei in the rare-earth region.}
// Chinese Physics C, Vol. {\bf 39}(4), 044101 (2015).

\bibitem{Usmanov010} Ph. N. Usmanov, A. A. Okhunov, U. S. Salikhbaev, A. I. Vdovin {\it Analysis of Electromagnetic Transitions in
Nuclei $^{176,178}Hf$}, Phys. Part. Nucl. Lett. {\bf 7}(3), 185--191 (2010).

\bibitem{Muhammad020} A. A Okhunov et al {\it Prediction of moment of inertia of rotating nuclei}, // Chinese Physics C Vol. {\bf 44}(11), 114107 (2020).

\bibitem{Hu93} J. Hu, F. Xu {\it Coupling between rotational and vibrational motions with the cranking Bohr-Mottelson Hamiltonian},
// Physical Review C, {\bf 48} (5) 2270 (1993).

\bibitem{Huang89} H. Huang, C. Wu, J. Zeng {\it Calculation of rotational spectra of well-deformed nuclei up to very high spins},
// Physical Review C, {\bf 39} (4) 1617 (1989).

\bibitem{Edmonds96} A.R.Edmonds {\it Angular momentum in quantum mechanics.}
// Princeton university press: {\bf 4}, (1996).

\bibitem{Landau013} L.D.Landau, E.M.Lifshizt {\it Quantum mechanics: non-relativistic theory.}
// Elsevier: Vol. {\bf 3}, (2013).

\bibitem{Pal83} M.K.Pal {\it Theory of nuclear structure.}
// Scientific and Academic Editions: (1983).

\bibitem{Greiner96} W.Greiner, J.A.Maruhn {\it Nuclear models.}
// Springer: (1996).

\bibitem{Http00} NuDat 2.8 {http://www.nndc.bnl.gov/nudat2/}

\end{thebibliography}

\end{document}